\documentclass[a4paper,11pt]{article}
\usepackage{jheppub}
\usepackage{amsmath,amssymb,bm}
\usepackage{array,multirow}
\usepackage{graphicx}
\hypersetup{colorlinks=true,citecolor=blue,linkcolor=blue,urlcolor=blue}
\newcommand{\ii}{\mathrm{i}}
\newcommand{\cM}{\mathcal M}
\newcommand{\cB}{\mathcal B}
\newcommand{\cG}{\mathcal G}
\newcommand{\Tr}{\operatorname{Tr}}
\newcommand{\Pf}{\operatorname{Pf}}
\title{Fractional Superfluid Vortices and Worldsheet Wess--Zumino--Witten Models in Two-Color QCD and $Sp(2N)$ Gauge Theories}

\author[a,b]{Muneto Nitta}
\affiliation[a]{Department of Physics \& Research and Education Center for Natural Sciences, Keio University, 4-1-1 Hiyoshi, Yokohama, Kanagawa 223-8521, Japan}
\affiliation[b]{International Institute for Sustainability with Knotted Chiral Meta Matter (WPI-SKCM$^2$), Hiroshima University, 1-3-1 Kagamiyama, Higashi-Hiroshima, Hiroshima 739-8531, Japan}

\abstract{
We study superfluid vortices in two-color QCD and more general pseudoreal gauge theories with even flavor number $N_f=2m$ at finite baryon density.  The antisymmetric diquark condensate admits fractional minimal vortices with baryon-phase winding $\nu=1/m$, corresponding to $1/m$ of the conventional circulation quantum, each localizing an $Sp(2)\simeq SU(2)$ core sector.  For vortices embedded in mutually orthogonal flavor planes, the leading-order chiral action factorizes exactly.  Off-diagonal flavor zero modes are non-normalizable at finite separation but become localized at coincidence, enhancing the normalizable sector from $[Sp(2)]^m$ to $Sp(2m)$.  For fundamental fermions of an $Sp(2N)$ gauge theory, the bulk Wess--Zumino--Witten (WZW) term descends compatibly: a cluster of $n$ coincident minimal vortices carries an $Sp(2n)_N$ WZW theory, while a flavor-symmetric vortex with positive integer baryon-phase winding $\nu$ carries $Sp(2m)_{N\nu}$.  At unit winding this yields the vortex map to $Sp(2m)_N$, with central charge $c=Nm(2m+1)/(N+m+1)$ at the WZW fixed point, and suggests a broader nonsupersymmetric four-dimensional/two-dimensional correspondence.
}

\begin{document}
\maketitle

\section{Introduction}

Gauge theories with fermions in pseudoreal representations display a pronounced flavor-number dependence already at zero density.  In two-color QCD (QC$_2$D) with fundamental Dirac fermions, the low-flavor theory is QCD-like and chirally broken, whereas increasing $N_f$ drives the system toward the conformal window.  This evolution has been investigated extensively on the lattice through the running coupling, mass anomalous dimension, finite-size scaling, and hadron spectrum \cite{Bursa2011,OhkiItou2010,Karavirta2012,HayakawaRunning2013,HayakawaSpectrum2013,Appelquist2014,Amato2016,Leino2017Nf8,Leino2018Nf6,LeinoSlope2018,Amato2018}.  The precise lower edge has historically been subtle, but later gradient-flow and spectral studies support an infrared fixed point for $N_f=8$ and place $N_f=6$ at or near the lower boundary of the conformal window \cite{Leino2017Nf8,Leino2018Nf6,LeinoSlope2018,Amato2018}.  Thus the zero-density infrared theory at general $N_f$ need not resemble ordinary low-flavor QC$_2$D.

A baryon chemical potential changes this infrared problem qualitatively.  It introduces an explicit scale and therefore removes the zero-density conformal fixed-point behavior.  In the finite-density diquark-condensed regime considered here, gauge-singlet pairing spontaneously breaks $U(1)_B$ and produces a baryonic superfluid \cite{Kogut1999,Kogut2000,SplittorffSonStephanov,SplittorffNLO,Adhikari2018}.  First-principles simulations have explored the resulting phase structure, equation of state, BEC--BCS evolution, and excitation spectrum \cite{Hands2006,IidaItouLee2020,Boz2020,Begun2022,IidaItou2022,IidaItou2024,ItouIida2024,ItouReview2025,ItouReview2026,IidaItou2026}.  Closely related pseudoreal theories, including $SU(2)\simeq Sp(2)$ and $Sp(4)$ gauge theories, are also widely studied as strongly coupled sectors in composite and dark-matter model building \cite{Hietanen2014,Hochberg2014,Hochberg2015,Bennett2018,Bennett2019,BennettSinglet2024,KulkarniEtAl,ZierlerStrongDM}.

Once a superfluid order parameter forms, its line defects provide a direct probe of the finite-density phase.  Their response to rotation, mutual forces, and reconnections underlies vortex lattices and turbulent dynamics across quantum fluids \cite{Sonin1987,SalomaaVolovik1987,Fetter2009,TsubotaKasamatsu2025}, with analogous defects playing an important role in neutron-star superfluids and in dense three-color QCD \cite{Link:2026exu,AlfordReview2008,EtoDenseReview2014,AlfordBaymFukushima2019}.  The internal structure is especially rich in the color--flavor-locked (CFL) phase, where non-Abelian semi-superfluid vortices carry localized $\mathbb{C}P^2$ orientational modes.  Their classical solutions, interactions, low-energy worldsheet theory, and the lifting of the orientational degeneracy by explicit flavor breaking have been studied in Refs.~\cite{BalachandranDigalMatsuura2006,NakanoNittaMatsuura2008,EtoNitta2009,EtoNakanoNitta2009,EtoNittaYamamoto2010}.  Quantum fluctuations of the relativistic $\mathbb{C}P^2$ worldsheet theory generate a strong scale and a mass gap, with kink excitations interpreted as confined monopoles \cite{GorskyShifmanYung2011,EtoNittaYamamoto2011}.  Supersymmetric gauge theories furnish another canonical realization of non-Abelian vortices, with localized $\mathbb{C}P^{N-1}$-type moduli and precisely controlled worldsheet dynamics \cite{HananyTong2003,AuzziEtAl2003,EtoEtAl2006,EtoHigherWinding2006,TongTASI2005,EtoReview2006,ShifmanYungRMP2007,ShifmanYungBook2009,TongReview2009}.  There as well, the relativistic worldsheet sigma model develops a dynamically generated mass scale and a massive spectrum, while its kinks encode confined monopoles and protected four-dimensional dynamics \cite{ShifmanYungMonopoles2004,HananyTongQuantum2004,Eto2011Mass}.  This relativistic mass-gap mechanism is not universal: for nonrelativistic vortices, type-II Nambu--Goldstone modes can remain gapless even quantum mechanically \cite{NittaUchinoVinci2014}, while supersymmetric vortices at specially tuned Argyres--Douglas points can instead flow to superconformal theories \cite{TongSC}.

Two-color QCD provides a qualitatively different relativistic outcome: the anomaly can protect a gapless vortex sector and drive it to a conformal WZW fixed point rather than the massive infrared theory of an ordinary sigma model.  For $N_f=2$, this mechanism was established in Ref.~\cite{Nitta:2026deq}.  In the chiral limit a unit baryonic vortex localizes exact normalizable $S^3\simeq SU(2)$ pion modes, while the bulk Wess--Zumino--Witten (WZW) term \cite{WessZumino1971,Witten1983}, when evaluated on the vortex background and integrated over the transverse directions, descends to a quantized WZW term for the localized worldsheet modes.  Closely related reductions of four-dimensional WZW couplings to lower-dimensional topological terms and effective interactions on chiral solitons have appeared for domain walls, chiral soliton lattices, and vortices in QCD chiral effective theories \cite{EtoNishimuraNitta2025PRL,EtoNishimuraNitta2024Rot,EtoNishimuraNitta2023Chain,AmariEtoNitta2025,CopingerEtoNittaQiu2026,QiuNitta2024,QiuNitta2025,QiuNitta2026}.  More generally, this bosonic WZW descent is related to, but conceptually distinct from, the anomaly-inflow mechanism on defects \cite{CallanHarvey1985,FukushimaImaki2018}.  The resulting $SU(2)$ principal chiral model has a level fixed by the microscopic anomaly: it is level one in QC$_2$D and level $N$ for fundamental fermions of an $Sp(2N)$ gauge theory.  The WZW fixed point is conformal and hence gapless, sharply contrasting with the dynamically massive relativistic vortex sigma models described above.  In this sense the bulk anomaly changes the qualitative infrared fate of the defect degrees of freedom.  This mechanism connects the bulk pseudoreal theory to exactly characterized two-dimensional WZW dynamics \cite{WittenBosonization1984,KZ1984,BPZ1984,DiFrancesco1997,PolyakovWiegmann1984,SchubringShifman2020,Coleman1973}.

In this work, we extend this structure to pseudoreal gauge theories with general even flavor number $N_f=2m$.  We keep the microscopic gauge representation fixed while increasing the number of flavors.  The essential new feature is that the antisymmetric diquark order parameter now has a nontrivial flavor orbit.  The topological generator of the superfluid vortex sector is no longer the vortex in which the full baryon phase winds by $2\pi$.  Instead, the minimal topological vortex combines a fractional $U(1)_B$ winding with a compensating flavor rotation.  Its circulation is $1/m$ of the conventional quantum, and the relevant integer topological charge is the winding of the Pfaffian.

A second, less obvious structure emerges when several minimal vortices are combined.  Minimal vortices can be embedded in mutually orthogonal symplectic flavor planes.  At leading order the corresponding multi-vortex ansatz factorizes exactly into $m$ copies of the $N_f=2$ problem, producing a no-force family.  The symmetry directions that mix different planes are exact zero directions of the static energy, but at finite separation they change the asymptotic winding plane and are therefore non-normalizable.  Exactly at coincidence their long-distance tail disappears, and the same directions become localized zero modes.  The dynamical moduli thus jump from $[Sp(2)]^m$ on the separated locus to $Sp(2m)$ at full coincidence.  This coincidence-induced enhancement is reminiscent of coincident D-branes, where stretched-string states become massless at coincidence and complete an enhanced non-Abelian gauge sector \cite{Witten1995Branes}.  Here, however, the off-diagonal directions are already exact static zero directions at finite separation and change instead from non-normalizable to localized.

Our main results are therefore threefold.  First, the vortex topology is measured by the Pfaffian winding $Q_{\rm Pf}=m\nu$, giving a fractional minimal circulation $\nu=1/m$.  Second, mutually orthogonal fractional vortices exhibit coincidence-induced localization of off-diagonal zero modes and a stratification of the normalizable moduli by cluster type.  Third, for fundamental fermions of an $Sp(2N)$ gauge theory the bulk WZW functional descends compatibly with this stratification: a single minimal vortex carries $Sp(2)_N$, a cluster of $n$ coincident minimal vortices carries $Sp(2n)_N$, and full coincidence yields $Sp(2m)_N$ with central charge $Nm(2m+1)/(N+m+1)$.  More generally, a flavor-symmetric vortex with positive integer baryon winding $\nu$ carries $Sp(2m)_{N\nu}$.

It is useful to place this last result in the broader context of symplectic WZW theories.  Their level-one structure has been studied through exceptional modular invariants and spinon/exclusion-statistics constructions, including explicit analyses of $Sp(4)_1$ and general $Sp(2n)_1$ models \cite{BouwknegtNahm1987,BouwknegtSchoutens1999}.  Higher-level symplectic WZW theories, fermionization, and symplectic level--rank structures have also been investigated \cite{NaculichRiggsSchnitzer1990,MlawerNaculichRiggsSchnitzer1991,Verstegen1991,BaeLee2021,OstrikRowellSun2020}.  Related symplectic current algebras arise in condensed-matter and statistical-mechanical settings, including sigma models with target $Sp(2n)/U(n)$ at $\theta=\pi$ \cite{Fendley2001Theta}, disordered $d$-wave superconductors and class-CI topological-superconductor surfaces \cite{AltlandSimonsZirnbauer2002,FosterYuzbashyan2012,GhorashiLiaoFoster2018}, and symplectic Kondo problems \cite{Kimura2021Kondo,LiKonigVayrynen2023,KonigTsvelik2023,RenKonigTsvelik2024,LotemEtAl2024}.  Symplectic global symmetry alone does not force an $Sp(2N)$ WZW infrared theory; a fundamental integrable spin chain with $Sp(2n)$ global symmetry provides an instructive counterexample \cite{Martins2002SpChain}.  Thus the $Sp(2m)_N$ theories selected here by anomaly descent belong to an established CFT family, but arise from a distinct four-dimensional defect mechanism.

For fundamental fermions of an $Sp(2N)$ gauge theory, the microscopic anomaly coefficient is $N$.  At unit flavor-symmetric winding, the result may therefore be summarized by the structural map
\begin{equation}
 \text{4d }Sp(2N)\text{ gauge theory with }N_f=2m
 \quad\xrightarrow{\;\text{vortex}\;}\quad
 \text{2d }Sp(2m)_N\text{ WZW theory}.
 \label{eq:4d2dmap}
\end{equation}
Equation~\eqref{eq:4d2dmap} is a consequence of the anomaly descent derived below, not a conjecture.  Whether this exact defect map extends to a deeper four-dimensional/two-dimensional correspondence is discussed in the Summary and Discussion.

The paper is organized as follows.  Section~\ref{sec:bulk} formulates the even-flavor chiral theory and identifies the vacuum orbit.  Section~\ref{sec:vortices} classifies flavor-symmetric integer-winding and fractional minimal vortices and derives their localized core modes.  Section~\ref{sec:multi} studies orthogonal multi-vortex configurations and the normalizability transition at coincidence.  Section~\ref{sec:wzw} derives the general WZW descent.  Section~\ref{sec:discussion} summarizes the results and discusses their CFT implications, possible deformations, and the conjectural four-dimensional/two-dimensional interpretation.  Appendix~\ref{app:odd} discusses odd flavor number, and Appendix~\ref{app:vector} derives the vector off-diagonal variations and their transverse norms for separated and coincident winding pairs.

\section{Symmetry, vacuum manifolds, and effective theory}
\label{sec:bulk}

We take an even number of flavors, $N_f=2m$, and use the convention that the flavor group $Sp(2m)$ has a $2m$-dimensional fundamental representation, so that $Sp(2)\simeq SU(2)$.  The microscopic pseudoreal gauge group is $Sp(2N)$, with fundamental fermions in its $2N$-dimensional representation.  Thus $N$ labels the gauge rank, while $m$ labels half the flavor number; QC$_2$D corresponds to $N=1$.  Even $N_f$ is required here for a full-rank antisymmetric flavor diquark condensate and its nonzero Pfaffian: with odd $N_f$ the condensate is necessarily rank deficient.  It is not imposed by a Witten global gauge anomaly~\cite{Witten1982Anomaly}.  Each Dirac fundamental flavor supplies two Weyl fundamentals, so their total number is $2N_f$ for any integer $N_f$; likewise an $SU(2)$ flavor subgroup acts on an even number, $2N$, of color copies.  We summarize the odd-flavor case in Appendix~\ref{app:odd}.
For flavor numbers for which the $\mu_B=0$ theory lies in or near the conformal window, the effective construction below is understood as a description of the finite-density diquark-condensed phase with the stated symmetry-breaking pattern; it is not a chiral expansion about a chirally broken zero-density vacuum.

When the zero-density theory is chirally broken, its Nambu--Goldstone fields take values in the pseudoreal-representation coset identified in Ref.~\cite{Kosower1984},
\begin{equation}
 \cM_0\equiv\frac{G_{\mu_B=0}}{H_{\mu_B=0}}
 =\frac{SU(4m)}{Sp(4m)}.
 \label{eq:zeroDensityCoset}
\end{equation}
The nonlinear chiral description used below is controlled when the enlarged $SU(4m)$ symmetry is spontaneously broken to $Sp(4m)$ at zero density and the chemical potential remains within the low-energy regime.  As in the original finite-density chiral effective theory \cite{Kogut2000}, typical momenta and $\mu_B$ must be small compared with the scale of non-Goldstone excitations; the gradients across the vortex core are of order $\mu_B$.  For flavor numbers whose zero-density theory lies in the conformal window, introducing finite density does not by itself establish the validity of the same nonlinear chiral field.  In that case our results are conditional on a diquark-condensed regime with the assumed effective field and symmetry-breaking pattern.

At finite baryon density, we select the gauge-singlet spin-zero diquark channel of the original finite-density effective theory \cite{Kogut1999,Kogut2000}.  Writing gauge indices as $a,b=1,\ldots,2N$ and flavor indices as $i,j=1,\ldots,2m$, the condensates may be written schematically as
\begin{equation}
 \Delta_L^{ij}\sim\Omega_{ab}\big\langle(q_L^{ia})^T Cq_L^{jb}\big\rangle,
 \qquad
 \Delta_R^{ij}\sim\Omega_{ab}\big\langle(q_R^{ia})^T Cq_R^{jb}\big\rangle.
\end{equation}
Here $q_{L,R}^{ia}$ are chiral quark fields, $C$ contracts the spinor indices, and $\Omega_{ab}=-\Omega_{ba}$ is the invariant antisymmetric tensor of the \emph{gauge} $Sp(2N)$ fundamental representation.  This gauge tensor is distinct from the flavor-space form $J_{2m}$ introduced below.  Fermi statistics and the two antisymmetric spinor and gauge contractions imply
\begin{equation}
 \Delta_{L,R}^T=-\Delta_{L,R}.
\end{equation}
At the Lie-group level the finite-density continuous symmetry is
\begin{equation}
 G_\mu\equiv SU(2m)_L\times SU(2m)_R\times U(1)_B,
\end{equation}
up to a finite quotient by common center elements.  We normalize the baryon charge of the diquark field to unity, as in Ref.~\cite{Nitta:2026deq}, so that
\begin{equation}
 \Delta_L\to e^{\ii\beta}g_L\Delta_Lg_L^T,
 \qquad
 \Delta_R\to e^{\ii\beta}g_R\Delta_Rg_R^T.
 \label{eq:DeltaTransformation}
\end{equation}
For even $N_f$ an antisymmetric matrix may have full rank.  We choose the reference symplectic form
\begin{equation}
 J_{2m}\equiv\operatorname{diag}(\epsilon,\ldots,\epsilon),
 \qquad
 \epsilon\equiv\begin{pmatrix}0&1\\-1&0\end{pmatrix},
 \qquad
 J_{2m}^2=-\bm 1_{2m}.
 \label{eq:J2m}
\end{equation}
The homogeneous superfluid vacuum can then be represented by
\begin{equation}
 \Delta_L=\Delta_R=J_{2m}.
 \label{eq:referencevac}
\end{equation}

The continuous stabilizer of Eq.~\eqref{eq:referencevac} is determined by
\begin{equation}
 g_{L,R}J_{2m}g_{L,R}^T=J_{2m},
 \qquad g_{L,R}\in Sp(2m),
\end{equation}
so that
\begin{equation}
 SU(2m)_L\times SU(2m)_R\times U(1)_B
 \longrightarrow Sp(2m)_L\times Sp(2m)_R.
 \label{eq:breaking}
\end{equation}
The corresponding finite-density vacuum orbit is the coset
\begin{equation}
 \cM_\mu\equiv\frac{G_\mu}{H_\mu}
 =\frac{SU(2m)_L\times SU(2m)_R\times U(1)_B}{H_\mu}
 \subset\cM_0,
 \qquad H_\mu^0=Sp(2m)_L\times Sp(2m)_R.
 \label{eq:finiteDensityCoset}
\end{equation}
Here $H_\mu$ is the full stabilizer of the reference vacuum, including finite baryon--flavor identifications, whereas $H_\mu^0$ is its connected component.  The inclusion in Eq.~\eqref{eq:finiteDensityCoset} identifies each finite-density vacuum with a configuration of the original chiral field.
The flavor part of this breaking is the standard even-flavor pattern of pseudoreal QCD-like theories \cite{Kogut2000,KanazawaWettigYamamoto2009,BraunerKolesova2019,LeeOhmoriTachikawa,Saito}.  At $m=1$ this additional flavor breaking is absent because $Sp(2)\simeq SU(2)$.

For the finite-density effective theory we retain the antisymmetric unitary field taking values in the zero-density chiral target $\cM_0$ of Eq.~\eqref{eq:zeroDensityCoset}; its homogeneous minima at finite $\mu_B$ lie on the subspace $\cM_\mu$ of Eq.~\eqref{eq:finiteDensityCoset}.  A convenient linearly transforming representative of the chiral field is a $4m\times4m$ matrix $\Sigma$.  In a basis adapted to the left- and right-handed diquark sectors we decompose it as
\begin{equation}
 \Sigma\equiv
 \begin{pmatrix}
  \Delta_L&\Phi\\
  -\Phi^T&-\Delta_R^*
 \end{pmatrix},
 \label{eq:SigmaBlock}
\end{equation}
where every block is $2m\times2m$.  The diagonal blocks are the antisymmetric diquark fields introduced above, while $\Phi$ is the baryon-neutral chiral (mesonic) block that interpolates between the left- and right-handed sectors; for $N_f=2$ it contains the $\sigma$ and pion fields of the original chiral effective theory \cite{Kogut1999,Kogut2000}.  The nonlinear sigma-model field obeys
\begin{equation}
 \Sigma^T=-\Sigma,
 \qquad
 \Sigma\Sigma^\dagger=\bm1_{4m},
 \qquad
 \det\Sigma=1,
 \qquad
 \Sigma\to G\Sigma G^T,
\end{equation}
with $G\in SU(4m)$.  The leading two-derivative chiral Lagrangian, with the baryon chemical potential introduced through a covariant derivative as in the original finite-density effective theory \cite{Kogut1999,Kogut2000}, is
\begin{equation}
 \mathcal L_2=\frac{f_\pi^2}{8}\Tr\!\left(D_\mu\Sigma D^\mu\Sigma^\dagger\right),
 \qquad
 D_0\Sigma\equiv\partial_0\Sigma-\ii\mu_B(B\Sigma+\Sigma B^T),
 \label{eq:L2}
\end{equation}
with
\begin{equation}
 B\equiv\frac12
 \begin{pmatrix}
  \bm1_{2m}&0\\0&-\bm1_{2m}
 \end{pmatrix}.
\end{equation}
Here $f_\pi$ is the low-energy decay constant, $\mu_B$ is the baryon chemical potential, $B$ is its Hermitian generator in the $4m$-dimensional chiral representation, and $D_i\Sigma\equiv\partial_i\Sigma$.  In particular, $B$ is distinct from the antisymmetric reference form $J_{2m}$ in the $2m$-dimensional flavor space.  This convention gives baryon charge one to the diquark blocks.

We impose $\Delta_L=\Delta_R\equiv\Delta$ only on the vortex vacuum at transverse spatial infinity, $r\to\infty$.  This is a boundary condition, not an equality imposed at finite $r$.  The reference profiles below satisfy the equality throughout space, but their allowed symmetry deformations need not do so away from infinity.  In particular, the axial off-diagonal transformations of Sec.~\ref{sec:multi} preserve the boundary condition for two winding pairs even when $\Delta_L\ne\Delta_R$ at finite $r$.  For vortex topology it is sufficient to follow this common asymptotic diquark order parameter $\Delta$.  Its vacuum manifold, namely the orbit of the reference condensate $J_{2m}$ under the faithful $SU(2m)\times U(1)_B$ action, is obtained by combining baryon and flavor transformations into $U(2m)$.  Since the stabilizer of $J_{2m}$ is $Sp(2m)$, this diquark vacuum manifold is
\begin{equation}
 \cM_\Delta\simeq\frac{U(2m)}{Sp(2m)}\subset\cM_\mu\subset\cM_0.
 \label{eq:vacmanifold}
\end{equation}
The first inclusion is the parity-symmetric vacuum subspace $\Delta_L=\Delta_R=\Delta$, represented by $\Sigma=\operatorname{diag}(\Delta,-\Delta^*)$ with $\Phi=0$.  It is imposed on the vortex boundary at infinity; a deformed vortex field at finite radius need not lie in this subspace, or even in the finite-density vacuum orbit $\cM_\mu$.
Since $Sp(2m)$ is simply connected,
\begin{equation}
 \pi_1(\cM_\Delta)\simeq\pi_1(U(2m))\simeq\mathbb Z.
 \label{eq:pi1}
\end{equation}
For $m=1$, $Sp(2)\simeq SU(2)$, and the vortex vacuum has only the baryon-number phonon as a bulk Nambu--Goldstone mode.  For $m>1$, each of the independent flavor breakings $SU(2m)_{L,R}\to Sp(2m)_{L,R}$ also produces bulk Nambu--Goldstone modes.  The additional massless bulk fields make the normalizability of flavor orientations a separate question from the vortex topology.  The integer winding is measured particularly naturally by the Pfaffian.

\section{Superfluid vortices and their effective worldsheet theories}
\label{sec:vortices}

\subsection{Pfaffian charge and circulation}

The Pfaffian of a $2m\times2m$ antisymmetric matrix is defined by
\begin{equation}
 \Pf\Delta\equiv\frac{1}{2^m m!}\epsilon^{i_1\cdots i_{2m}}
 \Delta_{i_1i_2}\cdots\Delta_{i_{2m-1}i_{2m}},
\end{equation}
with
\begin{equation}
 (\Pf\Delta)^2=\det\Delta,
 \qquad
 \Pf J_{2m}=1.
\end{equation}
Under $U\in U(2m)$, it transforms as
\begin{equation}
 \Pf(U\Delta U^T)=\det U\,\Pf\Delta.
\end{equation}
Thus $SU(2m)$ flavor rotations do not change the Pfaffian, whereas a baryon rotation gives
\begin{equation}
 \Pf(e^{\ii\beta}\Delta)=e^{\ii m\beta}\Pf\Delta.
\end{equation}
Defining the baryon-phase winding of a vortex at spatial infinity by
\begin{equation}
 \nu\equiv\frac{1}{2\pi}\oint d\beta,
\end{equation}
the integer that labels $\pi_1(\cM_\Delta)$ is the winding number of the Pfaffian itself.  We therefore define the Pfaffian topological charge
\begin{equation}
 Q_{\rm Pf}\equiv\frac{1}{2\pi\ii}\oint d\log\Pf\Delta=m\nu\in\mathbb Z.
 \label{eq:Qpf}
\end{equation}
In the long-distance region, the baryon phase gradient gives the superfluid momentum per unit diquark charge, while the temporal chemical-potential scale is $\mu_B$.  Thus, for $|\nabla\beta|\ll\mu_B$, the hydrodynamic flow velocity is $v_i\simeq\partial_i\beta/\mu_B$.  Its circulation around the vortex is
\begin{equation}
 \Gamma=\frac{2\pi}{\mu_B}\nu.
 \label{eq:circulation}
\end{equation}
Equation~\eqref{eq:Qpf} is the basic topological relation of the even-flavor theory.

\subsection{Singly quantized vortex}

For a straight vortex along the $z$ axis, let $(r,\theta)$ be polar coordinates in the transverse plane.  At spatial infinity, a singly quantized vortex has the asymptotic diquark field
\begin{equation}
 \Delta_{\rm sing}(r,\theta)\sim e^{\ii\theta}J_{2m},
 \qquad r\to\infty,
 \label{eq:Deltasing}
\end{equation}
where the symbol $\sim$ denotes the large-$r$ asymptotic form.  We call this vortex singly quantized because the full baryon phase winds once:
\begin{equation}
 \nu=1,
 \qquad
 \Gamma=\frac{2\pi}{\mu_B}.
\end{equation}
Its Pfaffian, however, winds $m$ times,
\begin{equation}
 \Pf\Delta_{\rm sing}\sim e^{\ii m\theta},
 \qquad Q_{\rm Pf}=m.
 \label{eq:singPf}
\end{equation}
Thus for $m>1$ the singly quantized vortex is not the minimal generator of $\pi_1$.

More generally, a flavor-symmetric vortex with positive integer baryon winding $\nu$ has
\begin{equation}
 \Delta_{\nu}(r,\theta)\sim e^{\ii\nu\theta}J_{2m},
 \qquad
 Q_{\rm Pf}=m\nu.
 \label{eq:integerwinding}
\end{equation}
The explicit profile and moduli analysis below is written for the unit case $\nu=1$; arbitrary integer winding will be reinstated in the WZW descent in Sec.~\ref{sec:wzw}.

A flavor-symmetric chiral-field ansatz realizing this asymptotic winding is
\begin{equation}
 \Sigma_{\rm sing}(r,\theta;g)=
 \begin{pmatrix}
  s e^{\ii\theta}J_{2m}&c g\\
  -c g^T&-s e^{-\ii\theta}J_{2m}
 \end{pmatrix},
 \qquad
 g\in Sp(2m),
 \label{eq:Sigmashing}
\end{equation}
where $s\equiv\sin\alpha(r)$ and $c\equiv\cos\alpha(r)$.  The radial profile obeys
\begin{equation}
 \alpha(0)=0,
 \qquad
 \alpha(\infty)=\frac{\pi}{2},
\end{equation}
so that the diquark condensate vanishes in the winding channel at the core and approaches the homogeneous superfluid vacuum at infinity.  The matrix $g$ specifies the orientation of the baryon-neutral chiral block $\Phi\equiv cg$ inside the core.  As the following energy functional shows, constant $g$ labels degenerate vortex solutions.  We will then show that its variations have finite transverse norm and therefore define localized worldsheet moduli.

\begin{figure}[ht]
 \centering
 \includegraphics[width=.58\textwidth]{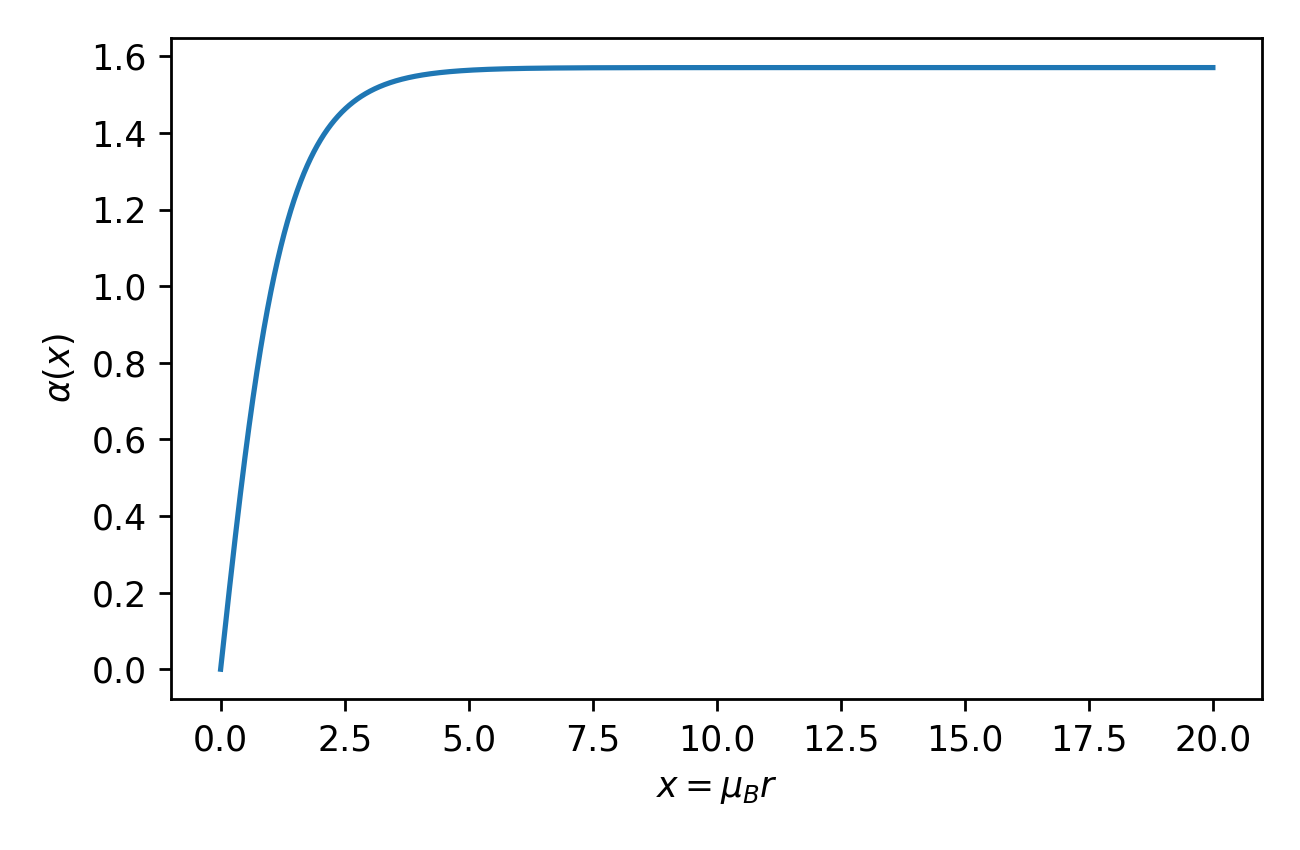}
 \caption{Universal chiral-limit radial profile of the active winding pair, plotted as a function of $x\equiv\mu_B r$.  The same solution appears for the $N_f=2$ vortex of Ref.~\cite{Nitta:2026deq} and for both the singly quantized and fractional minimal vortices considered here.  The chiral-core envelope is $\cos\alpha(x)$ and therefore decays exponentially in the homogeneous bulk.}
 \label{fig:profile}
\end{figure}

For constant $g$, substitution of Eq.~\eqref{eq:Sigmashing} into Eq.~\eqref{eq:L2} gives the static energy per unit length
\begin{equation}
 T_{\rm sing}=m\pi f_\pi^2\int_0^Rr\,dr
 \left[(\alpha')^2+\frac{\sin^2\alpha}{r^2}+\mu_B^2\cos^2\alpha\right],
 \label{eq:Tsing}
\end{equation}
where $R$ is a transverse infrared cutoff for the global-vortex tension.  The overall factor $m$ drops out of the Euler--Lagrange equation, leaving
\begin{equation}
 \alpha''+\frac{\alpha'}r-\frac{\sin\alpha\cos\alpha}{r^2}
 +\mu_B^2\sin\alpha\cos\alpha=0.
 \label{eq:profile}
\end{equation}
This is precisely the unit-vortex profile equation of the $N_f=2$ theory.  In terms of $x\equiv\mu_B r$, its regular numerical solution rises monotonically from $\alpha(0)=0$ to $\alpha(\infty)=\pi/2$, as shown in Fig.~\ref{fig:profile}.  Equivalently, $\beta(x)\equiv\pi/2-\alpha(x)$ decays exponentially at large $x$, $\beta\propto K_{\ii}(x)\sim e^{-x}/\sqrt{x}$, where $K_\nu$ denotes the modified Bessel function of the second kind.

Once this background solution is fixed, the symmetry orbit of the same $g$ already appearing in Eq.~\eqref{eq:Sigmashing} can be identified.  At the core, $c(0)=1$ and hence $\Phi(0)=g$.  The unbroken bulk $Sp(2m)_L\times Sp(2m)_R$ acts on the core orientation as $g\mapsto h_L g h_R^\dagger$.  Taking $g=\bm1_{2m}$ as a reference, its stabilizer is $Sp(2m)_V$, giving
\begin{equation}
 Sp(2m)_L\times Sp(2m)_R\longrightarrow Sp(2m)_V,
\end{equation}
and the corresponding internal family of degenerate vortex solutions is
\begin{equation}
 \cM_{\rm core}^{\rm sing}
 \equiv\frac{Sp(2m)_L\times Sp(2m)_R}{Sp(2m)_V}
 \simeq Sp(2m).
 \label{eq:coreSing}
\end{equation}
Here $Sp(2m)_V$ denotes the diagonal stabilizer of the reference orientation.  Equation~\eqref{eq:coreSing} describes the $g$ of Eq.~\eqref{eq:Sigmashing}, rather than an additional set of moduli.  Since its variations are multiplied by $c(r)$, which vanishes exponentially at large $r$, they are localized on the vortex.  These are classical vortex zero modes; the quantum $1+1$-dimensional theory does not require spontaneous symmetry breaking on the worldsheet.  They provide a non-Abelian analogue of the neutral-pion $U(1)$ core modulus of the charged-pion vortex in Ref.~\cite{QiuNitta2024}.

The normalization of the localized modes follows from the same profile equation.  The Pohozaev identity \cite{Nitta:2026deq} gives
\begin{equation}
 \int_0^\infty r\,dr\,\cos^2\alpha(r)=\frac{1}{2\mu_B^2}.
 \label{eq:Pohozaev}
\end{equation}
In the moduli approximation for soliton dynamics \cite{Manton1982,EtoEffective2006}, we promote the core orientation to a slowly varying field $g=g(t,z)$ on the vortex worldsheet.  Substituting this field into the bulk kinetic term and integrating over the transverse plane yields
\begin{equation}
 S_{{\rm kin},\rm sing}^{\rm ws}
 =\frac{\pi f_\pi^2}{4\mu_B^2}
 \int d^2x\,\Tr_{2m}(\partial_\gamma g^\dagger\partial^\gamma g),
 \qquad g\in Sp(2m).
 \label{eq:wsSingKin}
\end{equation}
Here $x^\gamma\equiv(t,z)$ are the coordinates along the straight vortex.  No additional overall factor $m$ appears in the worldsheet coefficient: the $m$ dependence is already contained in the fundamental trace $\Tr_{2m}$.

\subsection{Fractional minimal vortex}

We restrict this subsection and the multi-vortex construction below to the minimal sector $Q_{\rm Pf}=1$, in which the active symplectic pair winds once.  Repeated winding within the same pair introduces a separate winding multiplicity and is not needed for the coincidence physics studied here.

A canonical asymptotic representative of the generator $Q_{\rm Pf}=1$ is
\begin{equation}
 \Delta_{\rm min}(r,\theta)\sim\operatorname{diag}
 \bigl(e^{\ii\theta}\epsilon,\epsilon,\ldots,\epsilon\bigr),
 \qquad r\to\infty.
 \label{eq:Deltamin}
\end{equation}
Equivalently, its large-$r$ form may be decomposed as
\begin{equation}
 \Delta_{\rm min}(r,\theta)\sim e^{\ii\theta/m}h(\theta)J_{2m}h^T(\theta),
 \qquad r\to\infty,
 \label{eq:minDecomp}
\end{equation}
with
\begin{equation}
 h(\theta)\equiv\operatorname{diag}\left(
 e^{\ii(m-1)\theta/(2m)}\bm1_2,
 e^{-\ii\theta/(2m)}\bm1_{2m-2}\right)\in SU(2m).
\end{equation}
Therefore
\begin{equation}
 \nu=\frac1m,
 \qquad
 \Gamma=\frac{2\pi}{m\mu_B},
 \qquad
 Q_{\rm Pf}=1.
 \label{eq:minquantum}
\end{equation}
The vortex is fractional with respect to the conventional circulation quantum but minimal with respect to the actual $\pi_1$ generator.\footnote{The term ``fractional vortex'' also describes minimum-winding BPS vortices with fractional substructure in their transverse profiles \cite{Eto2009Fractional}.  Here it refers instead to the fractional baryon circulation $1/m$ fixed by the Pfaffian topology.}  This is the Pfaffian realization of the general principle that the minimal winding in a theory with nontrivial invariant polynomials is determined by the charge of the basic holomorphic invariant \cite{Eto2008Arbitrary,Eto2009SOUSp,Eto2009Quotients}.

A full chiral-field ansatz for a minimal vortex embedded in the first symplectic two-plane, which we call the active pair, is
\begin{align}
 \Delta
 &=\operatorname{diag}(s e^{\ii\theta}\epsilon,\epsilon,\ldots,\epsilon),
 \label{eq:minDelta}\\
 \Phi_{\rm min}
 &=\operatorname{diag}(c u,0_2,\ldots,0_2),
 \qquad u\in Sp(2)\simeq SU(2).
 \label{eq:minPhi}
\end{align}
Only the active pair carries transverse winding; all remaining symplectic pairs stay in the homogeneous diquark vacuum.  Consequently the static functional receives a nontrivial contribution only from that pair,
\begin{equation}
 T_{\rm min}=\pi f_\pi^2\int_0^Rr\,dr
 \left[(\alpha')^2+\frac{\sin^2\alpha}{r^2}+\mu_B^2\cos^2\alpha\right],
 \label{eq:Tmin}
\end{equation}
and the profile obeys the same Eq.~\eqref{eq:profile} shown in Fig.~\ref{fig:profile}.  Thus
\begin{equation}
 T_{\rm sing}=m T_{\rm min}.
 \label{eq:tensionrelation}
\end{equation}
The equality of radial profiles despite the different physical circulations reflects the compensating flavor rotation: the active pair itself winds through a full $2\pi$, whereas the net $U(1)_B$ winding is only $1/m$.

The internal structure of the minimal vortex now has two distinct pieces.  First, inside the active pair the core develops the same $S^3\simeq SU(2)$ chiral family as in the $N_f=2$ theory.  Writing
\begin{equation}
 u\equiv n^0\bm1_2+\ii n^a\tau_a,
 \qquad
 (n^0)^2+\bm n^2=1,
\end{equation}
where $\tau_a$ are the Pauli matrices, the local core symmetry breaking is
\begin{equation}
 Sp(2)_L\times Sp(2)_R\longrightarrow Sp(2)_V,
\end{equation}
with localized internal moduli
\begin{equation}
 \cM_{\rm core}^{\rm min}
 \equiv\frac{Sp(2)_L\times Sp(2)_R}{Sp(2)_V}
 \simeq Sp(2)\simeq SU(2)\simeq S^3.
 \label{eq:coreMin}
\end{equation}
Because the field $u$ is multiplied by the exponentially localized envelope $c(r)$, these $S^3$ directions have finite transverse norm and define genuine worldsheet fields.  In the moduli approximation \cite{Manton1982,EtoEffective2006}, promoting $u=u(t,z)$ while keeping the asymptotic winding-plane orientation fixed gives
\begin{equation}
 S_{{\rm kin},\rm min}^{\rm ws}
 =\frac{\pi f_\pi^2}{4\mu_B^2}
 \int d^2x\,\Tr_2(\partial_\gamma u^\dagger\partial^\gamma u),
 \qquad u\in Sp(2)\simeq SU(2).
 \label{eq:wsMinKin}
\end{equation}
Thus the localized target of a single minimal vortex is independent of $m$, while its embedding into the bulk flavor space is not.

Second, one may vary which symplectic two-plane in the $2m$-dimensional flavor space carries the winding.  Within the boundary condition $\Delta_L=\Delta_R$ at infinity, a common vector flavor rotation selects the winding plane of both chiralities.  The space of these winding-plane orientations is the quaternionic projective space
\begin{equation}
 \cB_{\rm orient}\equiv
 \frac{Sp(2m)}{Sp(2)\times Sp(2m-2)}
 =\mathbb H P^{m-1}.
 \label{eq:HPbase}
\end{equation}
Here $\mathbb H P^{m-1}$ is the space of possible symplectic two-planes selected as the active winding plane.  Unlike the localized $S^3$ core modes, changing this orientation changes the vortex field already at spatial infinity.  Global non-Abelian strings provide the general setting for such asymptotic orientational degrees of freedom \cite{NittaShiiki2008}.  More specifically, their transverse norm is infrared divergent, as in the orientation-dependent long-distance sector of global non-Abelian vortices \cite{NakanoNittaMatsuura2009,EtoNakanoNitta2009Global}.  These winding-plane orientations are consequently collective parameters of the global vortex but not dynamical worldsheet moduli.

The localized $S^3\simeq Sp(2)$ core modes and the non-normalizable winding-plane orientations $\mathbb H P^{m-1}$ fit together into a natural quaternionic fibration.  For $N_f=4$ ($m=2$), the full internal family is the quaternionic Hopf fibration
\begin{equation}
 S^3\longrightarrow S^7\longrightarrow S^4,
\end{equation}
with the group-theoretic identifications
\begin{equation}
 S^3\simeq\frac{Sp(2)\times Sp(2)}{Sp(2)},
 \qquad
 S^7\simeq\frac{Sp(4)}{Sp(2)},
 \qquad
 S^4\simeq\frac{Sp(4)}{Sp(2)\times Sp(2)}.
\end{equation}
For general $N_f=2m$ this becomes
\begin{equation}
 S^3\longrightarrow S^{4m-1}\longrightarrow\mathbb H P^{m-1},
 \label{eq:HopfGeneral}
\end{equation}
where
\begin{equation}
 \mathbb H P^{m-1}\simeq\frac{Sp(2m)}{Sp(2)\times Sp(2m-2)},
 \qquad
 S^{4m-1}\simeq\frac{Sp(2m)}{Sp(2m-2)}.
\end{equation}
The boundary-condition-preserving internal family in Eqs.~\eqref{eq:HopfGeneral} is therefore not the direct product $S^3\times\mathbb H P^{m-1}$: moving over the base also rotates the frame in which the $Sp(2)$ core variable is defined.  Physically, the $S^3$ fiber is localized by the core envelope $\cos\alpha(r)$, whereas the $\mathbb H P^{m-1}$ base changes the asymptotic winding orientation and is non-normalizable; hence only the fiber is a dynamical worldsheet target for an isolated minimal vortex.

\section{Orthogonal fractional vortices}
\label{sec:multi}

\subsection{Exact factorization and no-force family}

Place one minimal vortex in each of the $m$ mutually orthogonal symplectic flavor pairs.  Let $\bm x_\perp=(x,y)$ denote the transverse plane and let the $a$th vortex be centered at $\bm X_a$.  We define
\begin{equation}
 r_a\equiv|\bm x_\perp-\bm X_a|,
 \quad
 \theta_a\equiv\arg(\bm x_\perp-\bm X_a),
 \quad
 s_a\equiv\sin\alpha(r_a),
 \quad
 c_a\equiv\cos\alpha(r_a),
 \quad
 d_a\equiv s_a e^{\ii\theta_a}.
\end{equation}
The multi-vortex ansatz is
\begin{align}
 \Delta&=\operatorname{diag}(d_1\epsilon,\ldots,d_m\epsilon),
 \label{eq:multiDelta}\\
 \Phi&=\operatorname{diag}(c_1u_1,\ldots,c_m u_m),
 \qquad u_a\in Sp(2).
 \label{eq:multiPhi}
\end{align}
For this block-diagonal ansatz, each symplectic flavor pair forms exactly the same embedded $SU(4)/Sp(4)\simeq S^5$ chiral sector as the $N_f=2$ theory.  Since different pairs occupy disjoint matrix blocks, both the fundamental trace in the kinetic term and the chemical-potential contribution split into a sum over pairs, with no cross terms.  Consequently the leading-order (LO) action factorizes exactly into the actions of the $m$ embedded $N_f=2$ blocks,
\begin{equation}
 S_{\rm LO}=\sum_{a=1}^m S_{N_f=2}^{(a)}.
 \label{eq:factorization}
\end{equation}
For a common large-distance infrared cutoff,
\begin{equation}
 T_{\rm orth}(\bm X_1,\ldots,\bm X_m)=mT_{\rm min}
\end{equation}
is independent of all relative separations within this orthogonal ansatz.  Within this orthogonal truncation the positions are therefore exact classical flat directions.  This is the maximal-orientation analogue of the orientation-dependent interaction of global non-Abelian strings, for which the long-distance force vanishes at special relative orientations \cite{NittaShiiki2008,NakanoNittaMatsuura2009,EtoNakanoNitta2009Global}.

At generic separation the normalizable internal moduli consist simply of the product of the $m$ localized active-pair sectors,
\begin{equation}
 [Sp(2)]^m.
 \label{eq:separatedSym}
\end{equation}
The full bulk theory, however, has a larger flavor symmetry, and there are exact zero directions that mix distinct symplectic planes.  Whether those directions define worldsheet fields is a question of normalizability.

\subsection{Off-diagonal zero directions}

Consider two pairs, $a$ and $b$.  The remaining $m-2$ blocks play no role, so the problem closes in an $Sp(4)$ subspace.  By independent block-diagonal $Sp(2)_a\times Sp(2)_b$ transformations we may choose the reference point $u_a=u_b=\bm1_2$.  Define
\begin{equation}
 D_{ab}\equiv\operatorname{diag}(d_a\bm1_2,d_b\bm1_2),
 \qquad
 C_{ab}\equiv\operatorname{diag}(c_a\bm1_2,c_b\bm1_2),
\end{equation}
with $J_4\equiv\operatorname{diag}(\epsilon,\epsilon)$.  The four generators mixing the two symplectic planes span
\begin{equation}
 \mathfrak m_{ab}\equiv\mathfrak{sp}(4)-
 [\mathfrak{sp}(2)_a\oplus\mathfrak{sp}(2)_b],
\end{equation}
which is the tangent space of
\begin{equation}
 \frac{Sp(4)}{Sp(2)_a\times Sp(2)_b}\simeq S^4.
\end{equation}
Choose anti-Hermitian generators $T_A\in\mathfrak m_{ab}$, $T_A^\dagger=-T_A$, $A=1,\ldots,4$, normalized by
\begin{equation}
 \Tr_4(T_A^\dagger T_B)=\delta_{AB}.
\end{equation}

To follow these zero directions away from coincidence, we act with opposite left- and right-handed flavor rotations, i.e. with the axial combination
\begin{equation}
 g_L\equiv\exp\left(\frac{\eta^AT_A}{2}\right),
 \qquad
 g_R\equiv\exp\left(-\frac{\eta^AT_A}{2}\right).
\end{equation}
For two winding pairs, $d_a-d_b=O(R_{ab}/r)$ at infinity, so these axial rotations preserve the boundary equality $\Delta_L=\Delta_R$ as $r\to\infty$ even though they generally violate it at finite $r$.  This is why the boundary condition imposed in Sec.~\ref{sec:bulk} must not be extended to the entire transverse plane.  Using $T_AJ_4+J_4T_A^T=0$, the infinitesimal variations of the active two-pair block are
\begin{align}
 \delta\Delta_L&=\frac{\eta^A}{2}[T_A,D_{ab}]J_4,
 \label{eq:deltaDL}\\
 \delta\Delta_R&=-\frac{\eta^A}{2}[T_A,D_{ab}]J_4,
 \label{eq:deltaDR}\\
 \delta\Phi&=\frac{\eta^A}{2}\{T_A,C_{ab}\}.
 \label{eq:deltaPhi}
\end{align}
For an off-diagonal generator one may write schematically $T_A=\left(\begin{smallmatrix}0&X_A\\Y_A&0\end{smallmatrix}\right)$, so that
\begin{equation}
 [T_A,D_{ab}]\propto d_b-d_a,
 \qquad
 \{T_A,C_{ab}\}=(c_a+c_b)T_A.
\end{equation}
The diquark part therefore measures the difference of the two complex vortex profiles, whereas the chiral part measures the sum of their localized core envelopes.

The parameters $\eta^A$ are coordinates along these four off-diagonal symmetry directions.  Promoting them to slowly varying fields $\eta^A=\eta^A(t,z)$ gives the quadratic worldsheet kinetic term
\begin{equation}
 S_{ab,\rm off}^{(2)}
 =\frac{f_\pi^2}{16}\,\mathcal I_{ab}
 \sum_{A=1}^4\int d^2x_\parallel\,
 \partial_\gamma\eta^A\partial^\gamma\eta^A,
 \label{eq:Soff}
\end{equation}
where $d^2x_\parallel\equiv dt\,dz$.  The coefficient is the transverse norm
\begin{equation}
 \mathcal I_{ab}\equiv\int d^2x_\perp
 \left[|d_a-d_b|^2+(c_a+c_b)^2\right].
 \label{eq:Iab}
\end{equation}
This expression displays the normalizability transition directly.

At exact coincidence, $\bm X_a=\bm X_b$, one has $d_a=d_b=d$ and $c_a=c_b=c$.  The long-range diquark contribution vanishes pointwise, leaving
\begin{equation}
 \mathcal I_{ab}(0)=4\int d^2x_\perp\,c^2
 =8\pi\int_0^\infty r\,dr\,\cos^2\alpha
 =\frac{4\pi}{\mu_B^2},
 \label{eq:Izero}
\end{equation}
where Eq.~\eqref{eq:Pohozaev} was used.  Hence every off-diagonal generator obtains the finite kinetic coefficient
\begin{equation}
 \frac{f_\pi^2}{16}\mathcal I_{ab}(0)
 =\frac{\pi f_\pi^2}{4\mu_B^2},
 \label{eq:offCoeff}
\end{equation}
exactly equal to the normalization of the $Sp(2m)$ bi-invariant metric in Eq.~\eqref{eq:wsSingKin}.

At nonzero separation $R_{ab}\equiv|\bm X_a-\bm X_b|$, let $\varphi$ denote the polar angle of the distant observation point and $\varphi_{ab}$ the polar angle of the separation vector $\bm X_a-\bm X_b$.  The far field satisfies
\begin{equation}
 |d_a-d_b|^2\simeq(\theta_a-\theta_b)^2
 \simeq\frac{R_{ab}^2}{r^2}\sin^2(\varphi-\varphi_{ab}),
 \qquad r\gg R_{ab},\mu_B^{-1}.
\end{equation}
Therefore
\begin{equation}
 \mathcal I_{ab}(R_{ab})
 =\pi R_{ab}^2\log\frac{L}{\ell(R_{ab})}
 +\mathcal I_{\rm core}(R_{ab})+o(1),
 \qquad R_{ab}\neq0,
 \label{eq:Ilog}
\end{equation}
where $L$ is the transverse infrared cutoff, $\ell(R_{ab})$ is a finite matching scale set by the separation and core size, and $\mathcal I_{\rm core}(R_{ab})$ denotes the cutoff-independent contribution from the core and intermediate region.  The symbol $o(1)$ vanishes as $L\to\infty$.  The off-diagonal directions are exact flavor-symmetry directions, so the static potential has zero curvature along them.  Their fate is instead
\begin{align}
 R_{ab}=0 &: \quad \text{massless and normalizable},
 \label{eq:transition-coincident}\\
 R_{ab}\neq0 &: \quad \text{massless but non-normalizable}.
 \label{eq:transition}
\end{align}
Thus coincidence does not make a massive mode massless; it localizes a symmetry zero direction that already existed but had infinite norm.

For the same two winding pairs, an off-diagonal vector rotation ($g_L=g_R$) preserves $\Delta_L=\Delta_R$ at every radius.  When the vortices are separated, this symmetry direction has a logarithmically divergent transverse norm and does not define a dynamical worldsheet mode.  At exact coincidence it leaves the reference vortex invariant and becomes a stabilizer rather than an additional localized mode.  Appendix~\ref{app:vector} derives the vector variations and their norms explicitly.  By comparison, the axial rotations above preserve $\Delta_L=\Delta_R$ at infinity for separated winding pairs, although generally not at finite radius, and become localized at coincidence.  Table~\ref{tab:boundarymodes} summarizes these results and the directions mixing a winding pair with an unwound spectator.  In the last two rows ``unwound'' means a spectator flavor pair with $d_b=1$ and $c_b=0$.

\begin{table}[h!]
\caption{Off-diagonal symmetry directions classified by whether they preserve $\Delta_L=\Delta_R$ at transverse infinity.  The ``Boundary equality'' column concerns this asymptotic condition; normalizability is listed separately in the last column.  The core $Sp(2)$ mode of each winding pair is already localized and is not repeated here.}
\label{tab:boundarymodes}
\centering
\footnotesize
\renewcommand{\arraystretch}{1.15}
\setlength{\tabcolsep}{5pt}
\begin{tabular}{>{\raggedright\arraybackslash}p{0.25\textwidth}>{\raggedright\arraybackslash}p{0.16\textwidth}>{\raggedright\arraybackslash}p{0.26\textwidth}>{\raggedright\arraybackslash}p{0.22\textwidth}}
\hline
Pairs mixed & Action & Boundary equality & Transverse norm / role \\
\hline
\multirow{2}{=}{Separated winding pairs} & vector & preserved at all $r$ & $R_{ab}^{2}\log L$ \\
 & axial & preserved only as $r\to\infty$ & $R_{ab}^{2}\log L$ \\
\hline
\multirow{2}{=}{Coincident winding pairs} & vector & preserved at all $r$ & stabilizer, no zero mode \\
 & axial & preserved at all $r$ & finite, localized \\
\hline
\multirow{2}{=}{Winding / unwound pair} & vector & preserved at all $r$ & $L^2$, orientation \\
 & axial & violated at infinity & excluded by boundary condition \\
\hline
\end{tabular}

\end{table}

At coincidence, the axial off-diagonal directions supply the four localized modes that enlarge $Sp(2)_a\times Sp(2)_b$ to $Sp(4)$.  Mixing a winding pair with an unwound spectator by a vector transformation instead gives the non-normalizable orientation of Eq.~\eqref{eq:HPbase}; its axial counterpart changes $\Delta_L-\Delta_R$ already at infinity and lies outside the stated boundary conditions.  The asymptotic condition is independent of the $U(1)_A$ anomaly: these traceless symplectic rotations do not create a relative $U(1)_A$ phase or an anomaly-induced domain wall.

The separation-dependent normalizability in Eqs.~\eqref{eq:transition-coincident} and \eqref{eq:transition} has a close analogue in non-Abelian semilocal vortices \cite{ShifmanYung2006,Eto2007,ShifmanVinciYung2011}: orientational modes that are non-normalizable at generic nonzero size become normalizable on a special locus of moduli space \cite{Eto2007}.  Non-Abelian domain walls \cite{ShifmanYung2004Walls,EtoNittaOhashiTong2005,EtoFujimoriNittaOhashiSakai2008} provide a useful comparison for the coincidence phenomenon.  When separated constituent walls coincide, their internal symmetry is enhanced; in that case, however, the off-diagonal modes remain normalizable, and the reorganization is between Nambu--Goldstone and quasi-Nambu--Goldstone sectors rather than between non-normalizable and localized degrees of freedom \cite{Nitta2022Relations}.  A familiar string-theory analogue occurs for coincident D-branes: open strings stretched between separated branes are massive, whereas at coincidence their off-diagonal states become massless and complete an enhanced non-Abelian gauge sector \cite{Witten1995Branes}.  The present superfluid-vortex problem therefore combines the semilocal-vortex normalizability phenomenon with the coincidence enhancement familiar from non-Abelian domain walls, while differing from the D-brane mechanism because here the off-diagonal symmetry directions are already massless and change instead from non-normalizable to localized.

\subsection{Cluster stratification}

Suppose the $m$ minimal vortices are grouped into coincident clusters: vortices within the same cluster sit at the same transverse position, whereas different clusters remain separated.  If the cluster sizes are
\begin{equation}
 m=n_1+n_2+\cdots+n_p.
\end{equation}
Within each cluster the corresponding relative separations vanish and the off-diagonal directions become normalizable, whereas transformations mixing distinct clusters remain non-normalizable.  The localized symmetry is therefore
\begin{equation}
 \cG_{\rm loc}=\prod_{I=1}^p Sp(2n_I).
 \label{eq:clusterSym}
\end{equation}
The generic separated locus and full coincidence are the two extreme cases,
\begin{equation}
 [Sp(2)]^m
 \quad\longrightarrow\quad
 Sp(2m).
 \label{eq:enhancement}
\end{equation}
Equation~\eqref{eq:enhancement} is thus a statement about the dimension and group structure of the \emph{normalizable} dynamical moduli, not merely about the abstract bulk symmetry orbit.

\section{Wess--Zumino--Witten descent}
\label{sec:wzw}

The ungauged WZW action is defined for the full chiral field $\Sigma\in\cM_0=SU(4m)/Sp(4m)$, whose fifth homotopy group is $\pi_5(\cM_0)\simeq\mathbb Z$, rather than for the finite-density vacuum subspace $\cM_\mu$.  This distinction matters already at $m=1$: $\cM_0\simeq S^5$ has nontrivial $\pi_5$, whereas $\cM_\mu\simeq S^1$ does not.  A vortex explores the full chiral target because its core develops the mesonic block $\Phi$ and need not remain in $\cM_\mu$.  The transverse reduction below turns the five-form on $\cM_0$ into a three-form on the localized core target $Sp(2m)$, for which $\pi_3(Sp(2m))\simeq\mathbb Z$.

Let $M_5$ be a five-dimensional extension whose boundary is physical four-dimensional spacetime.  In the matrix normalization used above we write
\begin{equation}
 S_{\rm WZW}[\Sigma]
 =-\frac{\ii N}{240\pi^2}
 \int_{M_5}\Tr(\Sigma^\dagger d\Sigma)^5,
 \label{eq:WZWbulk}
\end{equation}
For the fundamental fermions of the $Sp(2N)$ gauge theory considered here, the microscopic flavor-anomaly coefficient is $N$ \cite{BraunerKolesova2019,LeeOhmoriTachikawa,Saito}; QC$_2$D corresponds to $N=1$.  The $N_f=2$ descent was derived explicitly in Ref.~\cite{Nitta:2026deq}; here we need its matrix generalization and its interplay with fractionalization.

\subsection{Flavor-symmetric integer vortices}

Equation~\eqref{eq:Sigmashing} describes the singly quantized, flavor-symmetric vortex with $\nu=1$.  Here ``flavor-symmetric'' means that every symplectic flavor pair winds in the same way.  For a positive integer baryon winding $\nu$, we replace $\theta\to\nu\theta$ and $\alpha(r)\to\alpha_\nu(r)$, with the same endpoint conditions $\alpha_\nu(0)=0$ and $\alpha_\nu(\infty)=\pi/2$.  The detailed radial profile depends on $\nu$, but the WZW level below depends only on the winding and the endpoint values.  To isolate the dependence on the core orientation, write
\begin{equation}
 \Sigma_{\nu}=G\Sigma_{0,\nu}G^T,
 \qquad
 G=\operatorname{diag}(g,\bm1_{2m}),
 \qquad
 \Sigma_{0,\nu}\equiv\Sigma_{\nu}\big|_{g=\bm1_{2m}},
\end{equation}
and define the Maurer--Cartan one-form on the three-dimensional WZW extension of the worldsheet by
\begin{equation}
 q\equiv g^{-1}dg\in\mathfrak{sp}(2m).
\end{equation}
The symplectic constraint is
\begin{equation}
 qJ_{2m}+J_{2m}q^T=0.
\end{equation}
Let $D_\perp^2$ denote the transverse disk spanned by $(r,\theta)$ and let $B_3$ be a three-manifold whose boundary is the $1+1$-dimensional vortex worldsheet.  For $A\equiv\Sigma^\dagger d\Sigma$, the component of the five-form on $D_\perp^2\times B_3$, containing one $d\alpha_\nu$, one $d\theta$, and three differentials along $B_3$, obeys the block-size-independent identity
\begin{equation}
 \left.\Tr A^5\right|_{D_\perp^2\times B_3}
 =40\ii\nu\sin\alpha_\nu\cos^3\alpha_\nu\,
 d\alpha_\nu\wedge d\theta\wedge\Tr_{2m}(q^3).
 \label{eq:traceid}
\end{equation}
After integrating the transverse disk,
\begin{equation}
 S_{\rm WZW}^{(\nu)}
 =\frac{\nu N}{12\pi}\int_{B_3}\Tr_{2m}(g^{-1}dg)^3
 =2\pi\nu N\,\Gamma_{Sp(2m)}[g],
 \label{eq:WZWsing}
\end{equation}
where
\begin{equation}
 \Gamma_{Sp(2m)}[g]
 \equiv\frac{1}{24\pi^2}\int_{B_3}\Tr_{2m}(g^{-1}dg)^3\quad(\mathrm{mod}\ \mathbb Z)
\end{equation}
is normalized so that the canonical one-block $Sp(2)\hookrightarrow Sp(2m)$ embedding has unit winding on a closed three-cycle.  For fixed worldsheet boundary data, two choices of extension over $B_3$ change $\Gamma_{Sp(2m)}[g]$ by an integer; an individual integral need not be an integer.  The integer coefficient $\nu N$ therefore makes the exponentiated Wess--Zumino action independent of the extension.  We use the notation $Sp(2m)_k$ to mean an $Sp(2m)$ group-valued worldsheet field with integer WZW level $k$.  This is the standard symplectic affine/WZW family.  At level one, $Sp(4)_1$ furnishes the classic realization of the $E_6$ modular invariant, while general $Sp(2n)_1$ theories admit spinon descriptions with generalized exclusion statistics \cite{BouwknegtNahm1987,BouwknegtSchoutens1999}.  Symplectic WZW theories at other ranks and levels also enter studies of fermionization and level--rank duality \cite{NaculichRiggsSchnitzer1990,MlawerNaculichRiggsSchnitzer1991,Verstegen1991,BaeLee2021,OstrikRowellSun2020}.  We will not use those detailed CFT results below; here the group and the integer level follow directly from the bulk vortex configuration and anomaly descent.  A flavor-symmetric vortex of positive integer winding $\nu$ therefore carries
\begin{equation}
 \text{localized WZW sector:}\qquad Sp(2m)_{\nu N}.
 \label{eq:SingWZW}
\end{equation}
For $\nu=1$ this is the singly quantized sector $Sp(2m)_N$ used in the coincidence analysis below.  The transverse reduction fixes the Wess--Zumino level, while the coefficient of the two-derivative worldsheet action is set by the vortex profile.  The principal chiral model with a level-$k$ Wess--Zumino term flows under the standard two-dimensional renormalization group toward the corresponding WZW fixed point \cite{SchubringShifman2020,WittenBosonization1984}, provided that the localized sector can be treated as an autonomous infrared theory.  At the $Sp(2n)_k$ fixed point, the Sugawara central charge is \cite{KZ1984,DiFrancesco1997}
\begin{equation}
 c\bigl[Sp(2n)_k\bigr]
 =\frac{k\,\dim Sp(2n)}{k+h^\vee_{Sp(2n)}}
 =\frac{k\,n(2n+1)}{k+n+1},
 \qquad h^\vee_{Sp(2n)}=n+1.
 \label{eq:symplecticCentralCharge}
\end{equation}
The symbol $h^\vee_{Sp(2n)}$ denotes the dual Coxeter number of the rank-$n$ symplectic Lie algebra.  Its value $n+1$, together with $\dim Sp(2n)=n(2n+1)$, gives the second equality.
Thus the flavor-symmetric winding-$\nu$ vortex has the central charge
\begin{equation}
 c_{\rm sym}(\nu)=\frac{N\nu\,m(2m+1)}{N\nu+m+1}.
 \label{eq:centralSymmetric}
\end{equation}
This statement concerns the flavor-symmetric higher-winding sector; its stability against splitting into lower-winding vortices is a separate dynamical question.  For $m=1$, Eq.~\eqref{eq:SingWZW} reproduces the $SU(2)_{N\nu}$ higher-winding hierarchy of Ref.~\cite{Nitta:2026deq}.

\subsection{Fractional vortices and coincidence}

For the minimal vortex only one symplectic pair depends on the transverse coordinates and on the localized internal field.  All other blocks are spectators.  Hence the five-form reduces exactly to the embedded $N_f=2$ block.  With $p\equiv u^{-1}du\in\mathfrak{sp}(2)\simeq\mathfrak{su}(2)$,
\begin{equation}
 \left.\Tr_{4m}(\Sigma_{\rm min}^\dagger d\Sigma_{\rm min})^5\right|_{D_\perp^2\times B_3}
 =40\ii\sin\alpha\cos^3\alpha\,
 d\alpha\wedge d\theta\wedge\Tr_2(p^3).
\end{equation}
Thus
\begin{equation}
 S_{\rm WZW}^{\rm min}
 =\frac{N}{12\pi}\int_{B_3}\Tr_2(u^{-1}du)^3
 =2\pi N\,\Gamma_{SU(2)}[u],
 \label{eq:WZWmin}
\end{equation}
and the minimal vortex carries
\begin{equation}
 \text{localized WZW sector:}\qquad Sp(2)_N\simeq SU(2)_N.
 \label{eq:MinWZW}
\end{equation}
Here and throughout the fractional-vortex construction the active-pair winding is one.  Although the physical baryon phase winds only by $2\pi/m$, the WZW level is $N$, not $N/m$.  The compensating flavor rotation makes the active symplectic pair itself execute a complete winding, and the fundamental trace simply reduces to the active $2\times2$ block.  Equivalently, the canonical one-block embedding $Sp(2)\hookrightarrow Sp(2m)$ has embedding index one.

For comparison, for the unit flavor-symmetric vortex $\nu=1$, the diagonal embedding
\begin{equation}
 j_{\rm diag}:SU(2)\hookrightarrow Sp(2m),
 \qquad
 u\mapsto\operatorname{diag}(u,\ldots,u),
\end{equation}
has embedding index $m$ because
\begin{equation}
 \Tr_{2m}\bigl[(j_{\rm diag}(u)^{-1}dj_{\rm diag}(u))^3\bigr]
 =m\Tr_2(u^{-1}du)^3.
\end{equation}
Therefore
\begin{equation}
 Sp(2m)_N\big|_{SU(2)_{\rm diag}}
 \Rightarrow SU(2)_{mN},
 \label{eq:diagembed}
\end{equation}
which mirrors the Pfaffian charge $Q_{\rm Pf}=m$ of the singly quantized vortex.

\begin{figure}[!htbp]
 \centering
 \includegraphics[width=0.72\textwidth]{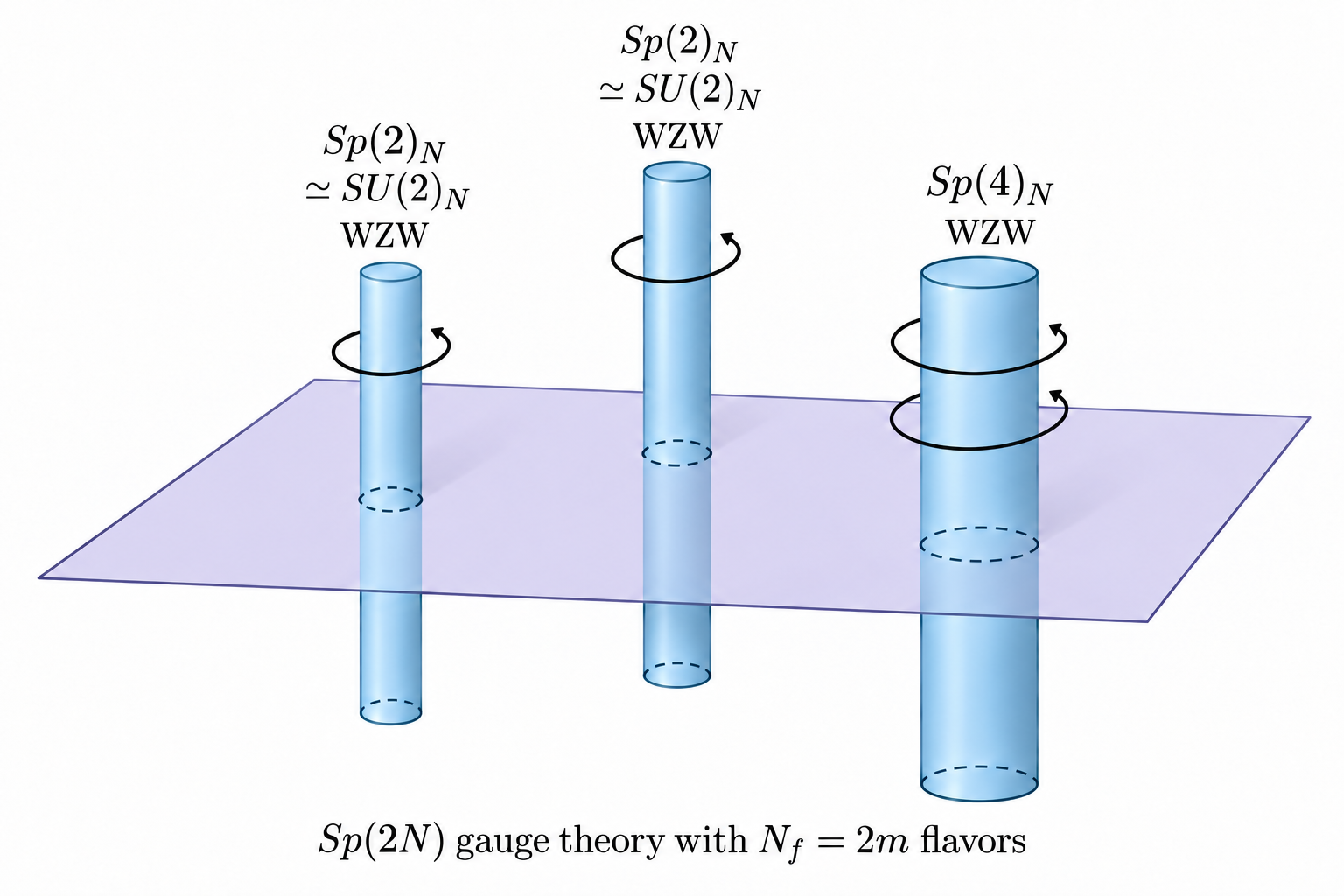}
 \caption{Schematic illustration of coincidence-induced worldsheet WZW enhancement in an $Sp(2N)$ gauge theory with $N_f=2m$ flavors.  The left and center tubes represent two separated fractional minimal vortices, each carrying a localized $Sp(2)_N\simeq SU(2)_N$ WZW sector.  Their exact coincidence produces the $Sp(4)_N$ sector shown on the right.  This two-vortex process is a representative step in the general cluster enhancement toward $Sp(2m)_N$; the three drawings compare separated and coincident configurations rather than depicting three vortices simultaneously.}
 \label{fig:WZWenhancement}
\end{figure}

On the block-diagonal subgroup $[Sp(2)]^m\subset Sp(2m)$,
\begin{equation}
 \Gamma_{Sp(2m)}[\operatorname{diag}(u_1,\ldots,u_m)]
 =\sum_{a=1}^m\Gamma_{SU(2)}[u_a].
 \label{eq:WZadditive}
\end{equation}
Hence $m$ separated orthogonal fractional vortices support factorized localized WZW sectors
\begin{equation}
 [Sp(2)_N]^m.
\end{equation}
At exact coincidence, the off-diagonal symmetry directions become normalizable and complete the full group manifold, giving
\begin{equation}
 [Sp(2)_N]^m
 \quad\xrightarrow{\ \text{exact coincidence}\ }\quad
 Sp(2m)_N.
 \label{eq:WZWenhancement}
\end{equation}
For fundamental fermions of an $Sp(2N)$ gauge theory, the representative two-vortex step $[Sp(2)_N]^2\to Sp(4)_N$ is illustrated in Fig.~\ref{fig:WZWenhancement}.

For a general cluster partition $m=\sum_I n_I$, the localized WZW sector is
\begin{equation}
 \prod_I Sp(2n_I)_N.
 \label{eq:clusterWZW}
\end{equation}
For a partition $m=\sum_I n_I$, let $c_{\rm sep}$ denote the combined central charge of the $m$ separated minimal-vortex sectors $[Sp(2)_N]^m$; $c_{\rm clusters}$, the sum over coincident clusters $Sp(2n_I)_N$; and $c_{\rm coinc}$, the central charge of the single $Sp(2m)_N$ sector at full coincidence.  Applying Eq.~\eqref{eq:symplecticCentralCharge} to these sectors gives
\begin{align}
 c_{\rm sep}&=\frac{3mN}{N+2},\nonumber\\
 c_{\rm clusters}&=\sum_I\frac{Nn_I(2n_I+1)}{N+n_I+1},\label{eq:centralClusters}\\
 c_{\rm coinc}&=\frac{Nm(2m+1)}{N+m+1}.\nonumber
\end{align}
The RG flow discussed above concerns the couplings within each worldsheet theory.
Thus the stratification of normalizable vortex moduli is lifted by anomaly descent to a corresponding stratification of quantized WZW theories; the same mechanism extends to arbitrary cluster partitions.

For QC$_2$D ($N=1$), the separated locus carries $[SU(2)_1]^m$ with $c_{\rm sep}=m$, whereas full coincidence carries $Sp(2m)_1$ with $c_{\rm coinc}=m(2m+1)/(m+2)$.  The enhancement of the worldsheet WZW sector therefore tracks the localization of the off-diagonal zero directions at coincidence.

\section{Summary and discussion}
\label{sec:discussion}

We have shown that increasing the number of flavors in a pseudoreal baryonic superfluid produces vortex topology and internal dynamics that do not arise at $N_f=2$.  At $N_f=2$ ($m=1$), the baryon-number phonon is the only bulk Nambu--Goldstone mode; for $m>1$, the flavor breaking $SU(2m)_{L,R}\to Sp(2m)_{L,R}$ yields additional bulk Nambu--Goldstone bosons.  For even $N_f=2m$, the vacuum orbit relevant to the antisymmetric diquark condensate is $U(2m)/Sp(2m)$, and the minimal vortex carries $1/m$ of the conventional baryon circulation quantum.  A conventional singly quantized vortex is a charge-$m$ configuration and admits a decomposition into $m$ fractional minimal vortices.

The fractional vortices carry two distinct kinds of internal structure.  Each vortex localizes the same $Sp(2)\simeq SU(2)$ core modes as the $N_f=2$ vortex, while the orientation of its winding symplectic plane lies in $\mathbb H P^{m-1}$ and is non-normalizable.  When minimal vortices occupy mutually orthogonal planes, the leading-order action factorizes and the vortices exert no force within that sector.  The off-diagonal flavor directions remain exact static zero directions, yet at finite separation they are not worldsheet fields because their norm diverges as $R^2\log L$.  Exactly at coincidence the asymptotic mismatch disappears and the same fields acquire a finite norm.  This gives coincidence-induced normalizable-moduli enhancement from $[Sp(2)]^m$ to $Sp(2m)$, and more generally a stratification by vortex clusters.

The anomaly follows the same geometry.  For fundamental fermions of an $Sp(2N)$ gauge theory, a fractional minimal vortex carries $Sp(2)_N$, a cluster of $n$ coincident minimal vortices carries $Sp(2n)_N$, and the full singly quantized vortex carries $Sp(2m)_N$.  More generally, a flavor-symmetric vortex with positive integer baryon winding $\nu$ carries $Sp(2m)_{N\nu}$.  Coincidence therefore changes not merely a classical moduli space but which quantized anomaly-carrying degrees of freedom are dynamically localized on the vortex.  The corresponding central charges are $c_{\rm sep}=3mN/(N+2)$ for separated minimal vortices and $c_{\rm coinc}=Nm(2m+1)/(N+m+1)$ at full coincidence, as derived in Sec.~\ref{sec:wzw}.  For QC$_2$D this produces the hierarchy from separated $[SU(2)_1]^m$ sectors to the coincident $Sp(2m)_1$ theory, while higher integer winding multiplies the level by $\nu$.

We now turn to the broader implications and open questions raised by these results.  The anomaly descent also singles out a suggestive color--flavor structure.  For fundamental fermions of a four-dimensional $Sp(2N)$ gauge theory, the anomaly coefficient is $N$: the flavor rank $m$ sets the affine symmetry on the vortex, while the color rank $N$ sets its WZW level.  Exchanging $N$ and $m$ in the same construction gives the pair
\begin{equation}
 \begin{array}{rcl}
  \text{4d }Sp(2N)\text{ gauge theory with }N_f=2m
  &\xrightarrow{\;\text{vortex}\;}& \text{2d }Sp(2m)_N\text{ WZW theory} \\[0.6em]
  && \Updownarrow\makebox[0pt][l]{\hspace{0.45em}\text{\scriptsize level--rank duality}} \\[0.6em]
  \text{4d }Sp(2m)\text{ gauge theory with }N_f=2N
  &\xrightarrow{\;\text{vortex}\;}& \text{2d }Sp(2N)_m\text{ WZW theory}
 \end{array}
 \label{eq:paired4d2d}
\end{equation}
Both horizontal maps in Eq.~\eqref{eq:paired4d2d} follow from vortex construction and anomaly descent.  The vertical link expresses level--rank duality between the resulting two-dimensional WZW theories; it does not assert a duality between their four-dimensional gauge theories.

This pairing relates primary labels, fusion coefficients, braid data, and modular structures without identifying the two local CFTs \cite{NaculichRiggsSchnitzer1990,MlawerNaculichRiggsSchnitzer1991,Verstegen1991,OstrikRowellSun2020}.  The same structure is encoded in the conformal embedding $\widehat{\mathfrak{sp}}(2m)_N\oplus\widehat{\mathfrak{sp}}(2N)_m\subset\widehat{\mathfrak{so}}(4mN)_1$, for which the central charges add to $2mN$ and the vector branches as the bifundamental $(\bm{2m},\bm{2N})$ \cite{Verstegen1991}.  Thus the two commuting affine factors are organized by precisely the flavor and color symplectic groups of the four-dimensional theory.  In particular, $m=N$ gives the level--rank self-dual vortex theory $Sp(2N)_N$.  This structure suggests, but does not establish, a microscopic interpretation in terms of non-Abelian bosonization.  Fermions carrying simultaneous color and flavor indices naturally admit commuting color and flavor current sectors, making it tempting to identify $Sp(2N)_m$ as a color sector and $Sp(2m)_N$ as its commuting flavor sector, with gauge dynamics removing the former while the latter survives on the vortex \cite{WittenBosonization1984}.  We have neither assumed a free-fermion ultraviolet worldsheet description nor established localized fermionic zero modes realizing this embedding, so this interpretation remains conjectural.  More generally, we conjecture that the exact anomaly map in Eq.~\eqref{eq:4d2dmap} may extend beyond the WZW level and global symmetry to a protected correspondence between vortex-bound four-dimensional operators or excitations and the primaries and fusion data of the two-dimensional $Sp(2m)_N$ WZW model.\footnote{For supersymmetric precedents, see Ref.~\cite{DoreyHollowoodTong1999} for exact 4d--2d BPS-spectrum matching; Refs.~\cite{HananyTong2003,AuzziEtAl2003,ShifmanYungMonopoles2004,HananyTongQuantum2004,ShifmanYungRMP2007,ShifmanYungBook2009,TongTASI2005,TongReview2009} for non-Abelian-vortex realizations and reviews; Refs.~\cite{DoreyHollowoodLee2011,ChenDoreyHollowoodLee2011} for exact 2d/4d relations from integrability and chiral-ring data; and Refs.~\cite{FujimoriKimuraNittaOhashi2012,FujimoriKimuraNittaOhashi2015} for vortex counting and vortex/instanton partition-function relations.}  The present setting is qualitatively different because there is no supersymmetry or BPS protection; the exact input is instead anomaly quantization.  We therefore regard the proposed operator-level correspondence as a conjectural nonsupersymmetric analogue.  The level--rank map of integrable representations provides a concrete first test of this conjecture.

The $m=1$ endpoint, including the higher-winding $SU(2)_{N\nu}$ sectors, connects the hierarchy to familiar one-dimensional quantum critical systems.  The $SU(2)_1$ WZW CFT governs the universal low-energy limit of the antiferromagnetic spin-$\tfrac12$ Heisenberg chain \cite{AffleckHaldane1987,ItoiMukaida1994}, and its finite-size spectrum and correlation functions give concrete universal data \cite{AffleckGepnerSchulzZiman1989}.  More generally, higher-level $SU(2)_k$ WZW theories occur at special critical points of higher-spin chains.  Half-odd-integer spin chains also obey Lieb--Schultz--Mattis-type constraints that obstruct a unique symmetric trivially gapped ground state \cite{LSM1961,Oshikawa2000,Tasaki2022}.  Here the microscopic obstruction has a different origin---the four-dimensional flavor anomaly descends to the vortex WZW term---but both mechanisms naturally enforce nontrivial infrared physics.  The present construction embeds this familiar $SU(2)$ endpoint into the broader symplectic hierarchy $Sp(2m)_N$ at unit winding and $Sp(2m)_{N\nu}$ for higher positive integer winding.

For $m>1$, vortices in the $N_f=2m$ theory are semilocal-like in the limited sense that a hypothetical gauging of $U(1)_B$ would leave the additional massless flavor fields in the bulk.  Our profiles are embedded solutions of the full chiral target.  Whether they admit extra size zero modes, and whether they are stable against deformations outside the embedded ansatz, have not been established.  This question is distinct from the symmetry-generated orientation modes whose norms were computed above.  In particular, the exact no-force result concerns the orthogonal block-diagonal ansatz at leading order.  In related $SO/USp$ gauge-theory vortices, local solutions and additional semilocal size sectors were distinguished in Refs.~\cite{Eto2008Arbitrary,Eto2009SOUSp}; the same possibility warrants a separate stability analysis here.

The internal modes studied here have finite transverse norms and, within the leading-order moduli approximation, form a localized worldsheet sector.  This differs from Kelvin modes, which move the vortex in the transverse plane and couple to long-range bulk fields.  Whether interactions with the gapless bulk flavor modes modify the infrared dynamics of the internal sector beyond this approximation remains an open question, analogous to the care required in defining effective worldsheet dynamics for semilocal vortices \cite{Eto2007}.

Several deformations deserve separate study.  Higher-derivative chiral operators will generically perturb the exact factorization and can lift the no-force family, although the topology and anomaly quantization are unchanged.  Finite quark mass is more consequential.  Already for $N_f=2$, the homogeneous condensate acquires a nonzero chiral component, $\cos\alpha_\infty=m_\pi^2/\mu_B^2$, so a uniform rotation of the core pion orientation changes the field all the way to infinity and is no longer normalizable \cite{Nitta:2026deq}.  For general even $N_f$ the same qualitative issue arises: explicit chiral-symmetry breaking deforms the asymptotic condensate and removes the exact chiral-limit moduli on which the WZW CFT construction rests.  The exact gapless $Sp(2m)_N$ statement should therefore be understood as a chiral-limit result.  Nevertheless, the vortex background can still support normalizable massive mesonic bound states, whose effective interactions and anomaly-induced couplings remain to be determined.

Pseudoreal $Sp(2N)$ gauge theories have been studied as confining dark sectors with composite dark pions and WZW interactions \cite{KulkarniEtAl,ZierlerStrongDM}.  Superfluid dark matter has also been proposed in a different microscopic setting, with light bosons condensing in galactic halos \cite{Berezhiani:2015bqa,Berezhiani:2025maf}.  If a pseudoreal dark sector carries a conserved dark baryon asymmetry and enters a diquark-condensed phase, it could support dark-baryon superfluid vortices.  The commonly studied $N_f=2$ case has unit minimal circulation, whereas $N_f=2m>2$ admits fractional minimal circulation $1/m$ and, in the chiral limit, the coincidence-dependent $Sp(2m)_N$ worldsheet sectors found here.  Whether a viable composite dark-matter model realizes this finite-density phase, and whether its vortices have observable consequences, remains to be investigated.

The collision and reconnection of the superfluid vortices studied here may be richer than in an ordinary $U(1)$ superfluid.  Already at $N_f=2$, the integer vortex carries a nontrivial localized $Sp(2)$ core sector.  For $m>1$, fractional vortices introduce flavor-dependent long-range interactions \cite{NakanoNittaMatsuura2009}, while coincidence of their constituents localizes additional internal directions.  Reconnection of ordinary superfluid vortices has long been studied analytically and numerically \cite{KoplikLevine1993,OgawaTsubotaHattori2002}.  For local non-Abelian vortices, the role of internal orientational moduli in collision and reconnection has been analyzed in Refs.~\cite{HashimotoTong2005,EtoHashimotoMarmoriniNittaOhashiVinci2007}.  The present global, anomaly-carrying vortices differ from both settings and deserve a dedicated dynamical study.

The worldsheet geometry opens further CFT directions.  A sufficiently large or stabilized closed vortex loop, or a vortex string compactified along its length, places the infrared theory on $\mathbb R_t\times S^1$.  Its finite-size spectrum should organize into conformal towers of $Sp(2m)_N$; universal finite-size corrections can then determine scaling dimensions and the central charge $c_{\rm coinc}$ from Eq.~\eqref{eq:symplecticCentralCharge} \cite{AffleckGepnerSchulzZiman1989,BloeteCardyNightingale1986}.  If a vortex is allowed to terminate on a physical boundary or interface, the worldsheet instead has a boundary and the appropriate infrared description is a boundary CFT, with endpoint conditions selecting boundary states and boundary operators \cite{Cardy1989Boundary}.  Likewise, a localized impurity coupled to the affine currents on the vortex would generate a Kondo-type boundary perturbation.  The standard CFT treatment of the Kondo problem \cite{AffleckLudwig1991Kondo}, together with modern symplectic Kondo constructions \cite{Kimura2021Kondo,LiKonigVayrynen2023,KonigTsvelik2023}, suggests that impurity physics on these vortices could provide a direct realization of $Sp(2m)_N$ boundary dynamics.

Most importantly, these predictions are accessible in principle to first-principles numerical tests.  Vortices and rotational responses have already been implemented in nonperturbative lattice-field simulations of quantum superfluids and gauge theories \cite{HayataYamamoto2015,Yamamoto2018,YamamotoHirono2013}.  In finite-density QC$_2$D one can therefore envisage constructing a quantized vortex and measuring flavor correlation functions along its core.  An isolated minimal vortex tests the $Sp(2)_1\simeq SU(2)_1$ sector, while coincident clusters at general even $N_f$ test the predicted enhancement to $Sp(2n)_1$; simulations of more general pseudoreal $Sp(2N)$ gauge theories would in addition test the anomaly-determined level $N$.  Finite-size scaling on a long periodic vortex or vortex loop could extract conformal dimensions and the central charge, while comparing separated and coincident vortices would directly probe the predicted normalizability enhancement.  Such calculations would test not only anomaly matching but also how far the proposed four-dimensional/two-dimensional correspondence extends into the operator and finite-size spectrum.  Because QC$_2$D is accessible to lattice simulation at finite baryon density, the vortex sector offers an unusually concrete first-principles route to realizing and testing an exactly characterized two-dimensional CFT localized inside a strongly coupled four-dimensional gauge theory.

\acknowledgments
The author thanks Taro Kimura for useful comments.
This work is supported in part by Japan Society for the Promotion of Science (JSPS) KAKENHI [Grants No.~JP22H01221 and JP23K22492] and the WPI program ``Sustainability with Knotted Chiral Meta Matter (WPI-SKCM$^2$)'' at Hiroshima University.

\appendix
\section{Odd flavor number}
\label{app:odd}

The restriction to even $N_f$ is essential to the construction developed in the main text.  For odd $N_f=2m+1$, every antisymmetric flavor matrix is necessarily rank deficient,
\begin{equation}
 \det\Delta=0,
 \qquad
 \operatorname{rank}\Delta\le 2m.
 \label{eq:odd-rank}
\end{equation}
At least one flavor direction therefore remains outside the diquark pairing.  A convenient reference condensate is
\begin{equation}
 \Delta_0\equiv
 \begin{pmatrix}
  J_{2m} & 0\\
  0 & 0
 \end{pmatrix},
 \label{eq:odd-reference}
\end{equation}
where the final flavor direction is unpaired.  This elementary rank deficiency is the first qualitative difference from even $N_f$: the full-rank orbit $U(2m)/Sp(2m)$ and its Pfaffian winding cannot simply be carried over to the odd theory.

There is also a continuous baryon--flavor compensation at the level of the order parameter.  Consider
\begin{equation}
 g(\beta)\equiv
 \operatorname{diag}\!\left(
 e^{-\ii\beta/2}\bm 1_{2m},
 e^{\ii m\beta}
 \right)\in SU(2m+1).
 \label{eq:odd-compensator}
\end{equation}
It acts on Eq.~\eqref{eq:odd-reference} as
\begin{equation}
 g(\beta)\Delta_0 g^T(\beta)=e^{-\ii\beta}\Delta_0,
 \qquad
 e^{\ii\beta}g(\beta)\Delta_0 g^T(\beta)=\Delta_0.
 \label{eq:odd-compensation}
\end{equation}
Thus a baryon phase rotation can be continuously compensated by a flavor rotation, in contrast with the essentially discrete baryon--flavor compensation of the full-rank even-flavor condensate.  This makes an embedded or semilocal-like interpretation natural.  It does \emph{not}, by itself, prove that the relevant vortex homotopy is trivial.  The faithful global group, the periodicity of $U(1)_B$, the baryon-charge normalization of the microscopic quarks, the quotient by the $SU(2m+1)$ center, and the relation between the left- and right-handed sectors all enter the exact homotopy problem.

A natural starting point is therefore to embed the even-flavor construction into the paired $2m$-dimensional subspace,
\begin{equation}
 \text{even-sector vortex}
 \ \hookrightarrow\
 \text{odd-}N_f\text{ theory}.
 \label{eq:odd-embedding}
\end{equation}
The smallest test case is $N_f=3$, for which
\begin{equation}
 \Delta_0^{(3)}=
 \begin{pmatrix}
  0&1&0\\
  -1&0&0\\
  0&0&0
 \end{pmatrix}.
 \label{eq:Nf3reference}
\end{equation}
The first two flavors form exactly the paired block of the $N_f=2$ theory, while the third flavor supplies an additional direction into which the configuration may spread.  Determining whether the embedded vortex is topologically stable, dynamically metastable, or can continuously unwind in the full target space requires the exact vacuum manifold and its global identifications.

The analogy with semilocal strings is useful precisely because classical family parameters need not define localized worldsheet degrees of freedom \cite{ShifmanYung2006,Eto2007}.  In the odd-flavor theory, extra flavor directions may generate power-law tails or size-like moduli, and the internal orientations visible inside the paired even-flavor block need not remain normalizable in the full theory.  At the same time the bulk theory still possesses its WZW functional, so a sufficiently long-lived embedded vortex could retain anomaly-induced interactions in a quasi-localized sector.  Whether an isolated WZW CFT survives, however, cannot be asserted without first settling the topology, stability, and normalizability of the odd-flavor vortex.  We therefore leave the odd case as a separate problem rather than extrapolating the even-flavor results beyond their controlled domain.

\section{Vector rotations of two winding pairs}
\label{app:vector}

For completeness we calculate the vector off-diagonal directions listed in Table~\ref{tab:boundarymodes}.  Work in the two-pair reference background $\Delta_L=\Delta_R=D_{ab}J_4$ and $\Phi=C_{ab}$ of Sec.~\ref{sec:multi}, and take the same normalized anti-Hermitian generators $T_A\in\mathfrak m_{ab}$ as in the axial calculation.  A vector transformation acts identically on the left and right,
\begin{equation}
 g_L=g_R=\exp\left(\frac{\xi^A T_A}{2}\right).
 \label{eq:vector-g}
\end{equation}
The infinitesimal variations are
\begin{align}
 \delta_V\Delta_L=\delta_V\Delta_R
 &=\frac{\xi^A}{2}[T_A,D_{ab}]J_4,
 \label{eq:vector-delta}\\
 \delta_V\Phi&=\frac{\xi^A}{2}[T_A,C_{ab}].
 \label{eq:vector-phi}
\end{align}
Here and below the repeated generator label $A$ is summed.  The equality of the two diquark blocks is thus preserved at every $r$, irrespective of the separation.  Since $T_A$ mixes the two blocks, the commutators in Eqs.~\eqref{eq:vector-delta} and \eqref{eq:vector-phi} are proportional to $d_b-d_a$ and $c_b-c_a$, respectively.  Promoting $\xi^A$ to slowly varying worldsheet fields gives, in the normalization of Eq.~\eqref{eq:Soff},
\begin{equation}
 S_{ab,V}^{(2)}=\frac{f_\pi^2}{16}\,\mathcal I_{ab}^{V}
 \sum_{A=1}^4\int d^2x_\parallel\,
 \partial_\gamma\xi^A\partial^\gamma\xi^A,
 \qquad
 \mathcal I_{ab}^{V}=\int d^2x_\perp
 \left[|d_a-d_b|^2+(c_a-c_b)^2\right].
 \label{eq:vector-norm}
\end{equation}
For $R_{ab}\ne0$, the $c_a-c_b$ term is localized near the cores, whereas $|d_a-d_b|^2$ has the same far-field tail as in Eq.~\eqref{eq:Ilog}.  Consequently
\begin{equation}
 \mathcal I_{ab}^{V}(R_{ab})
 =\pi R_{ab}^2\log\frac{L}{\ell(R_{ab})}
 +\mathcal I_{V,\mathrm{core}}(R_{ab})+o(1),
 \qquad R_{ab}\ne0.
 \label{eq:vector-log}
\end{equation}
The vector direction is therefore a static symmetry direction but is non-normalizable for separated vortices.  At exact coincidence, $d_a=d_b$ and $c_a=c_b$ pointwise, so both variations vanish and $\mathcal I_{ab}^{V}(0)=0$.  The vector $Sp(4)$ mixing is then part of the stabilizer of the coincident reference vortex, not an additional zero mode.  In contrast, the axial variation in Eq.~\eqref{eq:deltaPhi} contains $(c_a+c_b)T_A=2cT_A$ and produces the finite localized norm in Eq.~\eqref{eq:Izero}.  This distinguishes the vector and axial rows of Table~\ref{tab:boundarymodes} at coincidence.

\bibliographystyle{JHEP}
\bibliography{references}

\end{document}